# JPIC
# &
# How to make a PIC code

Hui-Chun Wu

E-mail: hcwu@lanl.gov

or huichunwu1@gmail.com

MS-E526, P-24,

Los Alamos National Laboratory,

Los Alamos, New Mexico 87545, USA

March 21, 2011

# Contents

# JPIC & How to make a PIC code

Hui-Chun Wu (武慧春)
E-mail: hcwu@lanl.gov or huichunwu1@gmail.com
*MS-E526, P-24, Los Alamos National Laboratory, New Mexico 87545, USA*

**Abstract**

Author developed the parallel fully kinetic particle-in-cell (PIC) code **JPIC** based on updated and advanced algorithms (e.g. numerical-dispersion-free electromagnetic field solver) for simulating laser plasma interactions. Basic technical points and hints of PIC programming and parallel programming by message passing interface (MPI) are reviewed. Most of contents come from Author's notes when writing up JPIC and experiences when using the code to solve different problems. Enough "how-to-do-it" information should help a new beginner to effectively build up his/her own PIC code. General advices on how to use a PIC code are also given.

## I. Introduction
### A. PIC method

The PIC method [1] simulates self-consistent interaction dynamics of particles (electrons and ions) and electromagnetic fields in a 1D, 2D or 3D mesh. PIC solves Maxwell's equations and Lorentz force equation (SI):

$$\nabla \times \mathbf{E} = -\frac{\partial \mathbf{B}}{\partial t} \quad (1)$$

$$\nabla \times \mathbf{B} = \mu_0 \varepsilon_0 \frac{\partial \mathbf{E}}{\partial t} + \mu_0 \mathbf{J} \quad (2)$$

$$\nabla \bullet \mathbf{E} = \frac{\rho}{\varepsilon_0} \quad (3)$$

$$\nabla \bullet \mathbf{B} = 0 \quad (4)$$

$$\frac{d\mathbf{P}}{dt} = q\left(\mathbf{E} + \mathbf{V} \times \mathbf{B}\right) \quad (5)$$

If one can make sure charge conservation

$$\frac{\partial \rho}{\partial t} + \nabla \bullet \mathbf{J} = 0, \quad (6)$$

by the use of a proper current calculation algorithm, e.g. charge-conservation scheme (CCS), Eqs (3,4) are not needed to be solved. Only Eqs. (1,2,5) are solved in JPIC. If one chooses a non-charge-conservation algorithm, Eq. (3) or Possion equation is really needed to be solved to correct the field. The latter needs global calculation in the whole mesh, which is not easy to be parallelized.

### B. PIC unit

First, it is convenient to normalize Eqs (1,2,3,5) to laser-related units:

$$\nabla \times \mathbf{E} = -\frac{\partial \mathbf{B}}{\partial t} \quad (7)$$

$$\nabla \times \mathbf{B} = \frac{\partial \mathbf{E}}{\partial t} + 2\pi \mathbf{J} \quad (8)$$

$$\nabla \bullet \mathbf{E} = 2\pi \rho \quad (9)$$

$$\frac{d\mathbf{P}}{dt} = 2\pi \frac{q}{M}\left(\mathbf{E} + \mathbf{V} \times \mathbf{B}\right) \quad (10)$$

Notice that symbols are kept same as before. Detailed normalizations are:

$$\frac{\mathbf{r}}{\lambda}, \frac{2\pi t}{\omega}, \frac{e\mathbf{E}}{m_e \omega c}, \frac{e\mathbf{B}}{m_e \omega}, \frac{\rho}{en_c}, \frac{\mathbf{J}}{en_c c}, \frac{\mathbf{P}}{Mc}, \frac{M}{m_e}, \frac{\mathbf{V}}{c}, \frac{q}{e} \quad (11)$$

where λ is the laser wavelength in vacuum, ω is the laser angular frequency, $n_c$ is the critical density, $m_e$ is the electron mass, e is the fundamental charge, c is the light speed in vacuum, and M can be the mass of electron or ion. Space and time are normalized to the laser wavelength and cycle respectively. Normalized laser amplitude $a_0 = eE_0/m_e \omega c$ is related to laser intensity by $I_0[\text{W/cm}^2] = 1.37 \times 10^{18} a_0^2 / \lambda_{\mu m}^2$ for linearly polarized laser, $\lambda_{\mu m}$ is the laser wavelength in micron. Now, the particle momentum is $\mathbf{P} = \gamma \mathbf{V}$, where the relativistic factor $\gamma = \sqrt{1+P^2} = 1/\sqrt{1-V^2}$.

The first advantage of normalization is making programming neat without disturbing specific units of physics quantities or constants. The second one is that one simulation done by use of normalized equations can be understood as infinite specific simulations. In PIC, laser wavelength is not specified and one can set it to an arbitrary value (λ=800nm, 8nm or 8m). Once the value of λ is chosen, all other physics quantities (plasma density, laser intensity etc.) are specified. For instance, one does a PIC simulation with parameters: $a_0$=1, $n_e$=0.5$n_c$, laser duration of 5 cycles and may find extremely nonlinear phenomena: large part of laser pulse energy is trapped inside the plasma, which forms a relativistic soliton [2]. If one sets λ=800nm, post-soliton size is about several microns, the initial laser intensity and plasma density is 2.14e18W/cm$^2$ and 8.67e20cm$^{-3}$. If one sets λ=8m, post-soliton size is about tens of meters, the initial laser intensity and plasma density is 2.14e4W/cm$^2$ and 8.67e6cm$^{-3}$. The former can be done in a table-top intense laser laboratory. The latter may be observed in the ionosphere or space if there are some strong microwave sources. This kind of scaling game is not trivial. In fact, one can share ideas and analyze physics mechanisms from one physics system/discipline to another physics system/discipline. Laboratory astrophysics [3] is trying to investigate space-scale problems in the laboratory plasma physics by scaling magnetohydrodynamics (MHD) equations.

**C. PIC coordinate**

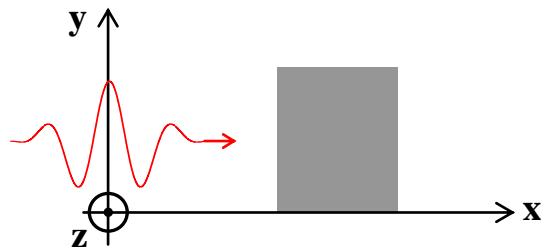

Figure 1. PIC coordinate.

Typically, PIC code uses x-axis as the longitudinal axis for laser propagation and y-z axes as transverse ones. In 1D PIC, space is shrunk to only x-axis and one cell is just a short line. In 2D PIC, simulation space is in x-y plane and one cell is square or rectangle. In 2D space, y-dir laser electric field is called as p-polarized and z-dir as s-polarized. In 3D PIC, simulation box is in real 3D space and one cell is cubic or cuboid.

**D. PIC code**

A basic PIC code generally contains several subroutines to read an input file, set up initial plasmas, inject a laser pulse, solve field equations, push particles, and output data. For C language users, they can be Input.c,

Plasma.c, Pulse.c, Field.c, Push.c, and Output.c. In the main routine Main.c, the flow chart can be

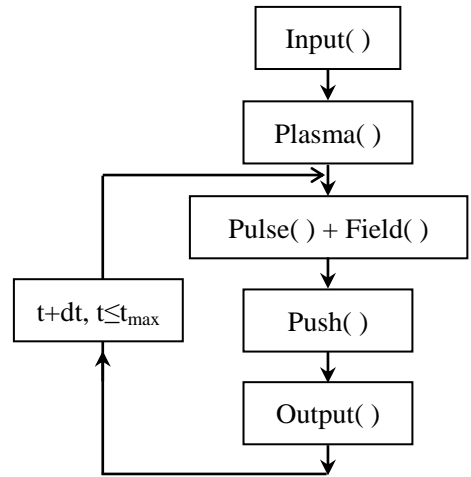

Figure 2. Flow chart of PIC.

Compiling Main.c generates an executable file main.exe. In the command line, input "main.exe input-deck" for running a job. One can find some ready routines for reading an ASCII input deck to make up the file Input.c.

In input deck, the necessary information is:

Simulation setup: total cell number in each dimension, cell number per wavelength in each dimension, particle number per cell.

Plasma: particle charge and mass, maximum density, length or cell number of uniform part, left ramp or right ramp, length of left vacuum, ramp profile (linear or exponential), electron and ion temperatures.

Laser: amplitude, shape, duration, polarization, injecting direction.

Output: maximum simulation time, time step of output snapshots, output path.

In Input.c, one also need allocate empty array or matrix for storing field, current, density, particle position, particle momentum, particle energy according to total cell number and calculated total particle number. For 100% safety, set them all to zero just after allocation.

In Plasma.c, first set momentum of each particle according to its temperature and make sure that momentum goes to zero for zero temperature. Second, set position of each particle and let them form a linear or exponential left ramp + flat + right ramp density profile within the simulation box. Since there is boundary condition for particles on boundaries, one should always leave a vacuum between initial particles and boundaries. How to arrange particles for a specific profile and set a particle boundary condition will be discussed below.

In Pulse.c, renew boundary fields on each time step, i.e. inject a laser pulse with choice shape and polarization. Pulse.c can be included into Field.c. In the beginning of Field.c, call Pulse( ) to update boundary fields. How to excite a laser pulse will be discussed below.

In Field.c, radiate the field by plasma currents and propagate them. In the meantime, one may be interested to accumulate field energy flux through the boundary for final energy diagnosis (input, reflected, and transmitted energy). Field solver and energy-related quantities in PIC unit will be discussed below.

In Push.c, interpolate field (only available on cell boundary or center) into particle itself (moves smoothly within the cell), use Lorentz force to push them, and obtain new position and momentum. In the meantime, calculate current and charge or number density to the mesh. How to do these will be discussed below.

In Output.c, output what one is interested: snapshots for field, density and current in space, and phase space information (position, momentum) and energy spectrum (some simple statistics) of all or selected particles. One can develop more elaborate diagnosis what one needs.

Before writing up a PIC code, one need sit down and derive and understand all the algorithms and technical details in each subroutine and get most of things ready for programming. According to essentiality, Field.c and JPIC                                                                                                                                    WUespecially Push.c are the heart/core of PIC. Any bug in them can be catastrophic.

All global variables used in above subroutines can be defined in a head file Variable.h.

**E. JPIC**

JPIC contains pure 1D and 2D versions: JPIC1d and JPIC2d. Both of them are parallel. Unlike commercial codes, where versatile abilities (oriented to various customers) demand many accessorial programming (much bigger than core parts of the code), JPIC is quite compact and easy to use.

The following table shows that line number of each subroutine in JPIC:

|        | Main.c | Variable.h | Input.c | Plasma.c | Pulse.c | **Field.c** | **Push.c** | Output.c | Total |
|--------|--------|------------|---------|----------|---------|-------------|------------|----------|-------|
| JPIC1d | 57     | 66         | 327     | 214      | 136     | **50**      | **308**    | 310      | 1468  |
| JPIC2d | 62     | 90         | 351     | 392      | 317     | **177**     | **552**    | 746      | 2687  |

Table 1. Line numbers in JPIC.

where comment and blank lines are also taken into account, and codes are not intensively optimized. In JPIC1d and JPIC2d, there are 50 and 155 lines of MPI parallel programming, respectively. Input decks of JPIC1d and JPIC2d are always smaller than 100 lines. Pulse.c contains two laser pulse injections with arbitrary polarizations, directions and six choice different shapes. Ouput.c contains enough diagnosis for most of problems. Plasma.c set a uniform plasma slab with both-side linear or exponential ramps. Codes will become a little bit bigger when there are special and advanced applications: e.g. atom ionization, moving window, and more species, which will be discussed in Section IV. Above all, own-made basic PIC code is really small!

## II. JPIC1d

Details on how to make a 1D PIC and related algorithms will be given below according to the flow chart in Fig. 2. JPIC1d has completely same algorithm as LPIC++ [4]. Here we go to more details and comments. The starting point is being familiar with layout of physics quantities on mesh/cell.

**A. Layout of physics quantities on cell**

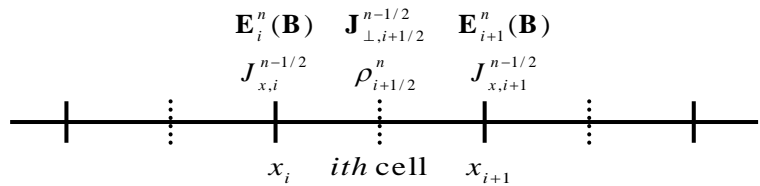

Figure 3. Layout of PIC1d: dashed vertical line: cell center, solid vertical line: cell boundary

All fields and longitudinal current are on cell boundaries and transverse currents and charge density are on cell centers. Boundaries of the ith cell are $x_i$ and $x_{i+1}$ and cell center is represented by i+1/2. n is non-negative integer and represents the time $n\Delta t$, half integer n+1/2 represent the middle time between $n\Delta t$ and $(n+1)\Delta t$.

**B. Field solving**
**1. Longitudinal field**

From Eq (8,9), one obtains

$$\frac{\partial E_x}{dt} = -2\pi J_x \qquad (12)$$

$$\frac{\partial E_x}{dx} = 2\pi\rho \tag{13}$$

Equations (12,13) are two ways to obtain $E_x$. Equations (7,8) implies that $B_x$ is always zero in 1D space.

Finite difference forms of Eqs. (12,13) are

$$E_{x,i}^{n+1} = E_{x,i}^{n} - 2\pi\Delta t J_{x,i}^{n+1/2} \tag{14}$$

$$E_{x,i+1}^{n} = E_{x,i}^{n} + 2\pi\Delta x \rho_{i+1/2}^{n} \tag{15}$$

where $\Delta t$ is time step and $\Delta x$ cell size and $\Delta t=\Delta x$ (light speed propagation). Equation (14) updates old $E_x$ to the next time step by the middle time current $J_x$. Equation (15) directly gets $E_x$ by collecting the charge density at the same time. When using CCS to get $J_x$, both methods give the same results. Equation (15) gives a way to check the correctness of the performance of CCS. If $E_x$ from Eq. (14) is correct, Eq. (14) is recommended.

## 2. P-polarized field ($E_y$)

$$\frac{\partial B_z}{dt} = -\frac{\partial E_y}{\partial x} \tag{16}$$

$$\frac{\partial E_y}{\partial t} = -\frac{\partial B_z}{dx} - 2\pi J_y \tag{17}$$

Rewrite them into

$$(\frac{\partial}{\partial t} + \frac{\partial}{dx})P_r = -\pi J_y \tag{18}$$

$$(\frac{\partial}{\partial t} - \frac{\partial}{dx})P_l = -\pi J_y \tag{19}$$

where

$$P_r = (E_y + B_z)/2, P_l = (E_y - B_z)/2 \tag{20}$$

are right-going and left-going waves, respectively. One has

$$E_y = P_r + P_l, B_z = P_r - P_l \tag{21}$$

Difference forms of Eqs. (18,19) are

$$P_{r,i+1}^{n+1} = P_{r,i}^{n} - \pi\Delta t J_{y,i+1/2}^{n+1/2} \tag{22}$$

$$P_{l,i}^{n+1} = P_{l,i+1}^{n} - \pi\Delta t J_{y,i+1/2}^{n+1/2} \tag{23}$$

Equation (22) propagates $P_r$ from left (i) to right (i+1) through the medium (time n+1/2 & space i+1/2) $J_y$ and Eq. (23) propagates $P_l$ from right to left.

## 3. S-polarized field ($E_z$)
Defining

$$S_r = (E_z - B_y)/2, S_l = (E_z + B_y)/2 \tag{24}$$

one has

$$E_z = S_l + S_r, B_y = S_l - S_r \tag{25}$$

Final difference forms for $S_{l,r}$ are

$$S_{r,i+1}^{n+1} = S_{r,i}^{n} - \pi \Delta t J_{z,i+1/2}^{n+1/2} \tag{26}$$

$$S_{l,i}^{n+1} = S_{l,i+1}^{n} - \pi \Delta t J_{z,i+1/2}^{n+1/2} \tag{27}$$

Checking with Eqs (22,23,26,27) in the vacuum ($J_{y,z}$=0), fields of a whole laser pulse are just moved by a grid $\Delta x$ on each time step $\Delta t$. In this way, even after infinite-steps calculation, laser pulse does not change the shape. This is a numerical dispersion free along x-dir (NDFX) field solver scheme. Conventional field solver based on classical time-domain time difference (FDTD) encounters from numerical dispersion: for a long-distance vacuum propagation, laser pulse will be stretched and chirped; more serious for few-cycle laser pulses, where absolute or carrier-envelope phase (CEP) changes quickly along x-dir. Complex CEP-stabilized schemes are already available for sub-terawatt few-cycle laser systems. PIC also needs a numerical CEP-stabilized scheme! Pukhov had proposed an upgraded FDTD-based NDFX scheme [5]. However, FDTD or FDTD-based scheme needs an absorption boundary to damp the reflected field, for instance the best choice, perfectly matched layers (PML) [6], which can obtain an amplitude reflection rate of <0.1% with a careful PML parameter choices. NDFX in Eqs (22,23,26,27) does not need any absorption boundary and reflection rate is null, which is beneficial to some simulations with a small signal output.

**C. Particle pushing**

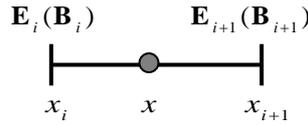

Figure 4. Fields are on ith cell boundaries ($x_i$&$x_{i+1}$) and gray ball is a particle with position x.

After updating fields by the above field solver, one needs linearly interpolate fields to each particle. Field on particle is

$$\mathbf{E} = \alpha \mathbf{E}_i + \beta \mathbf{E}_{i+1} \tag{28}$$

where $\alpha = (x_{i+1} - x)/\Delta x, \beta = (x - x_i)/\Delta x = 1 - \alpha$, and $x_i = i\Delta x, x_{i+1} = (i+1)\Delta x$. In real PIC programming, x is normalized for the second time to the cell size $\Delta x$. So one has

$$\alpha = i + 1 - x, \beta = x - i = 1 - \alpha \tag{29}$$

Where the symbol x is kept unchanged. One can know the cell index from particle position x by

$$i = (\text{int})x \tag{30}$$

Above all, obtain the cell index by Eq. (30) and then use Eqs. (28,29) to get fields on particles.

Then one can use Eq. (10) to push particle. Particle momentum is on half-integer time (n=1/2) and updated by integer-time fields. Equation (10) can be rewritten into

$$\frac{d\mathbf{P}}{dt} = 2\pi \frac{q}{M}(\mathbf{E} + \frac{\mathbf{P}}{\gamma} \times \mathbf{B}) \tag{31}$$

and it's difference form is

$$\frac{\mathbf{P}^{n+1/2} - \mathbf{P}^{n-1/2}}{\Delta t} = 2\pi \frac{q}{M}(\mathbf{E}^n + \frac{\mathbf{P}^{n+1/2} + \mathbf{P}^{n-1/2}}{2\gamma^n} \times \mathbf{B}^n), \tag{32}$$

Three steps to solve Eq. (32): half-acceleration, rotation and half-acceleration. Defining new quantities $\mathbf{P}^+, \mathbf{P}^-$ as

$$\mathbf{P}^- = \mathbf{P}^{n-1/2} - 2\pi \frac{q}{M} \mathbf{E}^n \frac{\Delta t}{2} \tag{33}$$

$$\mathbf{P}^{n+1/2} = \mathbf{P}^+ + 2\pi \frac{q}{M} \mathbf{E}^n \frac{\Delta t}{2} \tag{34}$$

Substitute Eqs (33,34) into Eq. (32), one gets

$$\mathbf{P}^+ - \mathbf{P}^- = 2\pi \frac{q}{M} \frac{\Delta t}{2\gamma^n} (\mathbf{P}^+ + \mathbf{P}^-) \times \mathbf{B}^n \tag{35}$$

where $\gamma^n = \sqrt{1 + P^{+2}} = \sqrt{1 + P^{-2}}$. Further define

$$\mathbf{T} = 2\pi \frac{q}{M} \frac{\Delta t}{2\gamma^n} \mathbf{B}^n \tag{36}$$

$$\mathbf{S} = \frac{2\mathbf{T}}{1 + T^2} \tag{37}$$

Then one has

$$\begin{cases} \mathbf{P}' = \mathbf{P}^- + \mathbf{P}^- \times \mathbf{T} \\ \mathbf{P}^+ = \mathbf{P}^- + \mathbf{P}' \times \mathbf{S} \end{cases} \tag{38}$$

where $\mathbf{P}'$ is a medium quantity. Equation (38) leads to

$$\mathbf{P}^+ = \begin{pmatrix} 1 - S_z T_z - S_y T_y & S_y T_x + S_z & S_z T_x - S_y \\ S_x T_y - S_z & 1 - S_x T_x - S_z T_z & S_z T_y + S_x \\ S_x T_z + S_y & S_y T_z - S_x & 1 - S_x T_x - S_y T_y \end{pmatrix} \mathbf{P}^- \tag{39}$$

Note that Eq. (39) is valid for a 2D or 3D case. Since $B_x=0$ in 1D case, i.e. $S_x=T_x=0$, Eq. (39) is simplified as

$$\mathbf{P}^+ = \begin{pmatrix} 1 - S_z T_z - S_y T_y & S_z & -S_y \\ -S_z & 1 - S_z T_z & S_z T_y \\ S_y & S_y T_z & 1 - S_y T_y \end{pmatrix} \mathbf{P}^- \tag{40}$$

Above all, 1st steps from $\mathbf{P}^{n-1/2}$ to $\mathbf{P}^-$ by Eq. (33), 2nd step from $\mathbf{P}^-$ to $\mathbf{P}^+$ by Eq. (39 or 40) and 3rd step from $\mathbf{P}^+$ to $\mathbf{P}^{n+1/2}$ by Eq. (34). The gamma factor and velocity is

$$\gamma^{n+1/2} = \sqrt{1 + (P^{n+1/2})^2} \tag{41}$$

$$\mathbf{V}^{n+1/2} = \mathbf{P}^{n+1/2} / \gamma^{n+1/2} \tag{42}$$

Displacement equation and its finite difference are

$$dx/dt = V_x \tag{43}$$

$$x^{n+1} = x^n + \Delta t V_x^{n+1/2} \tag{44}$$

In real PIC, x is normalized to the cell size $\Delta x$ and Eq. (44) becomes

$$x^{n+1} = x^n + V_x^{n+1/2} \tag{45}$$

## D. Current, charge density and plasma density calculation schemes

### 1. Current calculation

Check how to calculate currents in Ref. [4], where half of results are given. Here we list all the final results. Note that x below is already normalized to the cell size $\Delta x$. $x^n$ is old position (gray ball) and new position $x^{n+1}$ (orange ball) is obtained by Eq. (45). According to different locations of $x^n$ and $x^{n+1}$, there are six different cases.

1.1 $i \leq x^n < i+1/2$

1.1.1 $i-1/2 \leq x^{n+1} < i+1/2$

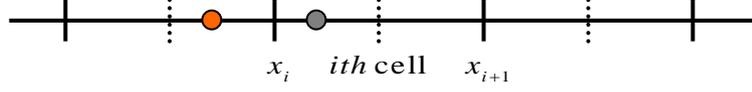

$$J_{x,j}^{n+1/2} = \rho_0 (x^{n+1} - x^n)$$

$$J_{y,j+1/2}^{n+1/2} = 0.5 \rho_0 V_y (1 + x^n + x^{n+1} - 2i)$$

$$J_{y,j-1/2}^{n+1/2} = \rho_0 V_y - J_{y,j+1/2}^{n+1/2}$$

1.1.2 $x^{n+1} < i-1/2$

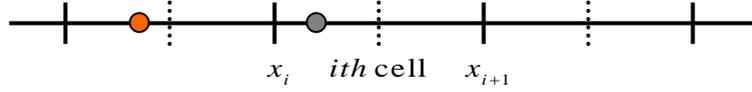

$$\varepsilon = \frac{x^n - i + 0.5}{x^n - x^{n+1}}$$

$$J_{x,j}^{n+1/2} = \rho_0 (-0.5 - (x^n - i))$$

$$J_{x,j-1}^{n+1/2} = \rho_0 (x^{n-1} - x^n) - J_{x,j}^{n+1/2}$$

$$J_{y,j+1/2}^{n+1/2} = 0.5 \varepsilon \rho_0 V_y (0.5 + x^n - i)$$

$$J_{y,j-3/2}^{n+1/2} = 0.5 (1-\varepsilon) \rho_0 V_y (-0.5 - (x^{n+1} - i))$$

$$J_{y,j-1/2}^{n+1/2} = \rho_0 V_y - J_{y,j+1/2}^{n+1/2} - J_{y,j-3/2}^{n+1/2}$$

1.1.3 $x^{n+1} \geq i+1/2$

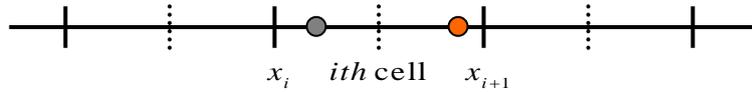

$$\varepsilon = \frac{x^n - i - 0.5}{x^n - x^{n+1}}$$

$$J_{x,j}^{n+1/2} = \rho_0 (0.5 - (x^n - i))$$

$$J_{x,j+1}^{n+1/2} = \rho_0 (x^{n-1} - x^n) - J_{x,j}^{n+1/2}$$

$$J_{y,j-1/2}^{n+1/2} = 0.5 \varepsilon \rho_0 V_y (0.5 - (x^n - i))$$

$$J_{y,j+3/2}^{n+1/2} = 0.5 (1-\varepsilon) \rho_0 V_y (x^{n+1} - i - 0.5)$$

$$J_{y,j+1/2}^{n+1/2} = \rho_0 V_y - J_{y,j-1/2}^{n+1/2} - J_{y,j+3/2}^{n+1/2}$$

1.2. $i+1/2 \leq x^n < i+1$

1.2.1 $i+1/2 \leq x^{n+1} < i+3/2$

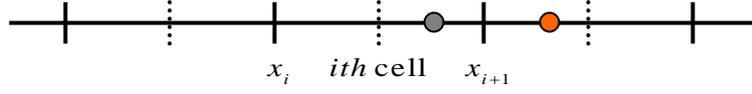

$$J_{x,j+1}^{n+1/2} = \rho_0(x^{n+1} - x^n)$$

$$J_{y,j+1/2}^{n+1/2} = 0.5\rho_0 V_y(3 + 2i - x^n - x^{n+1})$$

$$J_{y,j+3/2}^{n+1/2} = \rho_0 V_y - J_{y,j+1/2}^{n+1/2}$$

1.2.2 $x^{n+1} \geq i+3/2$

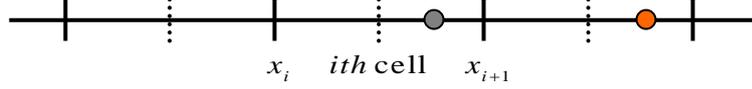

$$\varepsilon = \frac{x^n - i - 1.5}{x^n - x^{n+1}}$$

$$J_{x,j+1}^{n+1/2} = \rho_0(1.5 - (x^n - i))$$

$$J_{x,j+2}^{n+1/2} = \rho_0(x^{n-1} - x^n) - J_{x,j+1}^{n+1/2}$$

$$J_{y,j+1/2}^{n+1/2} = 0.5\varepsilon\rho_0 V_y(1.5 - (x^n - i))$$

$$J_{y,j+5/2}^{n+1/2} = 0.5(1-\varepsilon)\rho_0 V_y(x^{n+1} - i - 1.5)$$

$$J_{y,j+3/2}^{n+1/2} = \rho_0 V_y - J_{y,j+1/2}^{n+1/2} - J_{y,j+5/2}^{n+1/2}$$

1.2.3 $x^{n+1} < i+1/2$

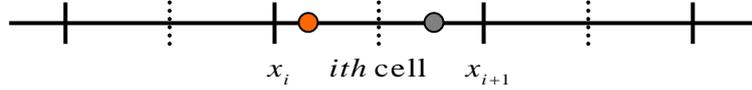

$$\varepsilon = \frac{x^n - i - 0.5}{x^n - x^{n+1}}$$

$$J_{x,j+1}^{n+1/2} = \rho_0(0.5 - (x^n - i))$$

$$J_{x,j}^{n+1/2} = \rho_0(x^{n-1} - x^n) - J_{x,j+1}^{n+1/2}$$

$$J_{y,j+3/2}^{n+1/2} = 0.5\varepsilon\rho_0 V_y(x^n - i - 0.5)$$

$$J_{y,j-1/2}^{n+1/2} = 0.5(1-\varepsilon)\rho_0 V_y(0.5 - (x^{n+1} - i))$$

$$J_{y,j+1/2}^{n+1/2} = \rho_0 V_y - J_{y,j-1/2}^{n+1/2} - J_{y,j+3/2}^{n+1/2}$$

$J_z$ is same as $J_y$ by simply replacing $V_y$ by $V_z$. Relevant coefficients in each case can simplify programming. $\rho_0$ is charge density of each macro-particle

$$\rho_0 = \frac{qn_e}{N_{cell}} \qquad (46)$$

where $n_e$ is the peak plasma density in $n_c$, particle charge q in e, and particle number per cell $N_{cell}$. JPIC always takes same total particle number for both electron and ion. So, $N_{cell}$ and abs($\rho_0$) are same for both of them.

**2. Charge density calculation**

To get charge density ρ at cell center at new time (n+1), there are two different cases:

**2.1** $(int)x^{n+1} \leq x^{n+1} < (int)x^{n+1} + 1/2$

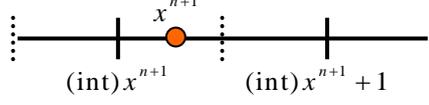

$$\rho^{n+1}_{(int)x^{n+1}+1/2} = \rho_0(x^{n+1} - (int)x^{n+1} + 0.5)$$

$$\rho^{n+1}_{(int)x^{n+1}-1/2} = \rho_0 - \rho^{n+1}_{(int)x^{n+1}+1/2}$$

**2.2** $(int)x^{n+1} + 1/2 \leq x^{n+1} < (int)x^{n+1} + 1$

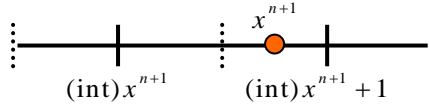

$$\rho^{n+1}_{(int)x^{n+1}+1/2} = \rho_0(1.5 - x^{n+1} + (int)x^{n+1})$$

$$\rho^{n+1}_{(int)x^{n+1}+3/2} = \rho_0 - \rho^{n+1}_{(int)x^{n+1}+1/2}$$

Charge density ρ is only used in Eq. (15) as a second way to solve $E_x$.

**3. Plasma density calculation**

Calculate plasma density is only for diagnosis. We define it on cell boundaries

$$n^{n+1}_{(int)x^{n+1}+1} = n_0(x^{n+1} - (int)x^{n+1})$$

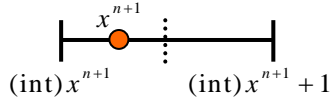

$$n^{n+1}_{(int)x^{n+1}} = n_0 - n^{n+1}_{(int)x^{n+1}+1}$$

where $n_0$ is plasma density of each macro-particle

$$n_0 = \frac{n_{e,i}}{N_{cell}} \tag{47}$$

where $n_{e,i}$ is electron or ion density.

**E. Make Input.c**

From input deck, one gets total cell number: cn, particle number per cell: pn_per_cell, cell number of left vaccum region: cn_vac, cell number of uniform plasma slab: cn_flat, and cell numbers of left or right ramp: cn_lramp and cm_rramp, cell number per wavelength: cn_per_wl.

Allocate array for quantities on cell boundary: $J_x$[0:cn], others are $E_{x,y,z}$, $B_{y,z}$, $P_{l,r}$, $S_{l,r}$,

Allocate array for quantities on cell center: $J_{y,z}$[0:cn-1] and ρ[0:cn-1];

Allocate matrix for plasma density on cell boundary: den[0:1][0:cn], where index [0][:] and [1][:] represent electron and ion, respectively.

Calculate total particle (electron or ion) number: for uniform plasma slab, it is pn=pn_per_cell*cn_flat; for a linear or exponential ramp, please refer to the next section.

Allocate matrix for position, momentum and gamma factor of particles: X[0:1][0:pn-1], $P_{x,y,z}$[0:1][0:pn-1] and gamma[0:1][0:pn-1].

Set all of above quantities to zero.

Calculate Debye length in cell size by

$$\frac{DL}{\Delta x} = \frac{\sqrt{T_e / m_e c^2}}{2\pi \sqrt{n_e / n_c}} \times \text{cn\_per\_wl} \tag{48}$$

which typically is demanded to be close to or larger than one.

    Set initial time: t=0.

    Set cell size: dx=1/cn_per_wl.

    Set time step: dt=dx.

    Set charge density and number density of single micro-particle according to Eqs. (46,47).

    Set time steps (from input deck) to periodically output.

## F. Make Plasma.c
### 1. Set initial momentum of particles according to given temperatures

$T_e$ and $T_i$ can be different. Refer to Ref. [1] for how to do it.

### 2. Set position of particles
### 2.1 Uniform plasma slab

Calculate separation of particles sp=1/pn_per_cell;

Set position X[:][i]=cn_vac+sp*(i+0.5).

### 2.2 Linear ramp

Set a linear ramp from $n^0$ to $n^1$. Particle index as a function of position is $i(x) = \xi(ax^2 + bx)$, $i \in [0, np-1]$, its derivative gives density profile $di/dx \propto n(x) = 2ax + b$, $n(0) = n^0$, $n(x_L) = n^1$, $x_L$ is plasma length. One has $a = (n^1 - n^0)/2x_L$ and $b = n^0$. In $i(x)$, there are $i(0) = 0$ and $i(x_L) = np-1$, so $\xi = \dfrac{np-1}{x_L(n^1+n^0)/2}$.

Substitute a,b and $\xi$ into i(x), and solve x as

$$x = \frac{\sqrt{n^0 n^0 (np-1)^2 + (n^1 + n^0)(n^1 - n^0)i(np-1)} - n^0(np-1)}{(n^1 - n^0)(np-1)} x_L \tag{49}$$

where $n^1/n^0 > 1$ or $n^1/n^0 < 1$. The final density profile is

$$n(x) = (n^1 - n^0)x/x_L + n^0 \tag{50}$$

Total particle number is proportional to $\int_0^{x_L} n(x)dx = (n^1 + n^0)x_L/2$.

### 2.3 Exponential ramp
### 2.3.1 Rising ramp

Particle index as a function of position is $i(x) = af_e n_0 \exp[(x-L)/f] + b$, $i \in [0, np-1]$, L is the ramp length. $di/dx \propto n(x) = n_0 \exp[(x-L)/f_e]$, where $n(0) = n_0 \exp(-L/f_e)$, $n(L) = n_0$. $i(x)$ has $i(0) = 0$, $i(L) = np - 1$. One gets

$$a = \frac{np-1}{f_e n_0[1-\exp(-L/f_e)]}, b = -\frac{np-1}{1-\exp(-L/f_e)}\exp(-\frac{L}{f_e}), i(x) = \frac{np-1}{\exp(-L/f_e)-1}[\exp(x/f_e)-1], \text{ and}$$

$$x = f_e \ln\{\frac{i}{np-1}[\exp(L/f_e)-1]+1\} \tag{51}$$

The final density profile is

$$n(x) = n_0 \exp(\frac{x-L}{f_e}) \tag{52}$$

Total particle number is proportional to $\int_0^L n(x)dx = n_0 f_e[1-\exp(-L/f_e)]$.

**2.3.2 Dropping ramp**

Particle index as a function of position is $i(x) = af_e n_0 \exp(-x/f_e) + b$, $i \in [0, np-1]$, L is ramp length. $di/dx \propto n(x) = n_0 \exp(-x/f_e)$, where $n(0) = n_0$, $n(L) = n_0 \exp(-L/f_e)$. $i(x)$ has $i(0) = 0$, $i(L) = np-1$.

One gets

$$a = \frac{np-1}{f_e n_0[\exp(-L/f)-1]}, b = \frac{np-1}{1-\exp(-L/f_e)}, \quad i(x) = \frac{np-1}{\exp(-L/f)-1}[\exp(-x/f)-1], \text{ and}$$

$$x = -f_e \ln\{\frac{i}{np-1}[\exp(-L/f_e)-1]+1\} \tag{53}$$

$$n(x) = n_0 \exp(-x/f_e) \tag{54}$$

Total particle number is proportional to $\int_0^L n(x)dx = n_0 f_e[1-\exp(-L/f_e)]$.

**2.4 Arbitrary functional profile**

For an arbitrary density profile, there may be no analytic solution for position as a function of particle index. One can use a random-number generator to easily build any kind of density profile. For a profile function $n(x) = n_0 f(x)$ in region $x \in [0, L]$, where $n_0 = \max(n(x))$ or $\max(f(x)) = 1$: there is np_per_cell particles on the position where $n = n_0$; first generate an uniform particle distribution with peak density $n_0$ and the length L; then using a random number generator ran() to produce a floating-point number within (0,1); when $\text{ran}() \leq f(x)$, this particle is really allocated, otherwise jump over. Random-number based particle initialization demands a large particle number per cell. Otherwise there are big noises on density profiles.

**2.5 Variable particle weight**

Above methods are based on the same particle weight, i.e. Eq. (46) is same for all particles. The other way to build an arbitrary plasma profile is changing the particle weight. Particles can be uniformly distributed in the space. The plasma density is proportional to the particle weight. One need open an additional array to storing each particle relative weight, like W[0:1][0:pn-1]. When making diagnoses for energy spectrum, particle weights need to be taken into accounts. This approach may be good for load balance of parallelization.

### G. Make Pulse.c, Field.c and Push.c

In Pulse.c, p-polarized right-going waves from left boundary can be injected by $P_{r,0}^n = a_0 f(n\Delta t)\cos(2\pi n\Delta t)$, where f is the pulse shape function and n is the number of time steps. And left-going waves from right boundary can be $P_{r,cn}^n = a_0 f(n\Delta t)\cos(2\pi n\Delta t)$. The same is for s-polarized. For circular polarization, right-going wave can be $P_{r,0}^n = a_0 f(n\Delta t)\cos(2\pi n\Delta t)$ and $S_{r,0}^n = a_0 f(n\Delta t)\sin(2\pi n\Delta t)$.

At the beginning of Field.c, call Pulse() to renew boundary fields. Use Eqs. (14,15,22,23,26,27) in Section C to renew fields from n to n+1.

Follow the procedure in Sections C&D to make the Push.c file. If some particles move to the x-dir boundary, they are reflected with a thermal velocity.

Outputting total field and particle energies is important for checking convergence of simulations. Here we give field and particle energy in PIC unit:

Energy density of electromagnetic field is $w_{eb} = (\varepsilon_0 E^2 + B^2/\mu_0)/2$. Electromagnetic energy in one cell length dx and arbitrary transverse area A is $W_{eb,cell} = w_{eb} A dx$. Kinetic energy of particles in one cell length and arbitrary transverse area A is $W_{p,cell} = [m_e c^2 n_{e0}\sum_e(\gamma_e - 1) + m_i c^2 n_{i0}\sum_i(\gamma_i - 1)]A dx$, where $n_{e0,i0} = n_{e,i}/N_{cell}$ is number density of single macro-particle (electron or ion).

Energy is normalized to $m_e c^2 n_c A dx$ and other quantities to PIC unit:

$$W_{p,cell} = n_{e0}\sum_e(\gamma_e - 1) + m_i n_{i0}\sum_i(\gamma_i - 1), \quad (55)$$

$$W_{eb,cell} = (E^2 + B^2)/2 \quad (56)$$

Make a sum $W_{total} = \sum_{all-cells}(W_{eb,cell} + W_{p,cell})$ to get total energy in the whole simulation box.

Electromagnetic energy flux is $\mathbf{S} = \mathbf{E}\times\mathbf{B}/\mu_0$. Incident/transmitted laser energy from boundaries is $W_{eb} = \int \mathbf{S}\cdot\mathbf{A}\, dt$. Normalize it to $m_e c^2 n_c A dx$ and PIC unit:

$$W_{eb} = \int |\mathbf{E}\times\mathbf{B}| = \sum_n (E_y^n B_z^n + E_z^n B_y^n) \quad (57)$$

When laser is totally contained in simulation box, all-cells sum of Eq. (56) and Eq. (57) will give the same results.

### H. Oblique incidence

Check Ref. [4] for how to implement oblique incidence by a Lorentz transformation. Moving frame has a velocity $\beta=\sin(\theta)$ along y-dir and relativistic factor $\gamma=1/\cos(\theta)$. In the moving frame: laser frequency is $\omega^M=\omega^L/\gamma$, laser becomes normal incidence, and plasma flows along –y direction with β with a increased density $n^M=\gamma n^L$. Here superscripts L and M represent lab and moving frames, respectively. PIC is done in this moving frame and PIC unit is related to $\omega^M$ light. Below, we list Lorentz transformation from moving frame to lab frame for all physics quantities, which is useful for diagnosis. In normalized forms, left Quantity[L] is normalized to $\omega^L$ related PIC unit in the lab frame and right Quantity[M] is normalized to $\omega^M$ related PIC unit in the moving frame.

**1. Electromagnetic field**

Dimensional forms:

$$E_x^L = \gamma(E_x^M - \beta B_z^M) \qquad E_y^L = E_y^M \qquad E_z^L = \gamma(E_z^M + \beta B_x^M)$$

$$B_x^L = \gamma(B_x^M + \beta E_z^M) \qquad B_y^L = B_y^M \qquad B_z^L = \gamma(B_z^M - \beta E_x^M) \qquad B_x^M = 0$$

Normalize them to PIC unit in both lab and moving frames:

$$E_x^L = E_x^M - \beta B_z^M \qquad E_y^L = E_y^M / \gamma \qquad E_z^L = E_z^M$$

$$B_x^L = \beta E_z^M \qquad B_y^L = B_y^M / \gamma \qquad B_z^L = B_z^M - \beta E_x^M$$

**2. Density and current**

Dimensional forms for total charge density and currents:

$$\rho^L = \gamma(\rho^M + \beta J_y^M / c)$$

$$J_x^L = J_x^M, \quad J_z^L = J_z^M, \quad J_y^L = \gamma(\beta c \rho^M + J_y^M)$$

**2.1 Number density**

Dimensional forms: $\quad -en_e^L = \gamma(-en_e^M + \beta J_{e,y}^M / c) \quad Zen_i^L = \gamma(Zen_i^M + \beta J_{i,y}^M / c)$

Normalized forms: $\quad n_e^L = (n_e^M - \beta J_{e,y}^M)/\gamma \qquad n_i^L = (n_i^M + \beta J_{i,y}^M / Z)/\gamma$

**2.2 Y-dir current**

Dimensional forms: $\quad J_{e,y}^L = \gamma(-\beta c e n_e^M + J_{e,y}^M) \qquad J_{i,y}^L = \gamma(\beta c Z n_i^M + J_{i,y}^M)$

Normalized forms: $\quad J_{e,y}^L = (-\beta n_e^M + J_{e,y}^M)/\gamma \qquad J_{i,y}^L = (\beta Z n_i^M + J_{i,y}^M)/\gamma$

**2.2 XZ-dir current**

Dimensional forms: $\quad J_{e,x}^L = J_{e,x}^M \qquad J_{i,x}^L = J_{i,x}^M \qquad J_{e,z}^L = J_{e,z}^M \qquad J_{i,z}^L = J_{i,z}^M$

Normalized forms: $\quad J_{e,x}^L = J_{e,x}^M / \gamma^2 \qquad J_{i,x}^L = J_{i,x}^M / \gamma^2 \qquad J_{e,z}^L = J_{e,z}^M / \gamma^2 \qquad J_{i,z}^L = J_{i,z}^M / \gamma^2$

**3. Momentum and energy** (normalized)

$$\gamma^L = \gamma(\gamma^M + \beta P_y^M)$$

$$P_x^L = P_x^M, \quad P_z^L = P_z^M, \quad P_y^L = \gamma(P_y^M + \beta \gamma^M)$$

**I. Parallelization**

JPIC1d is parallelized by MPI, where Author only parallelizes particles. This strategy is simpler and based on the fact: total memory of 1D simulation is typically small and pushing particles consumes most of simulation time. First initialize all the particles in the primary rank/processor. Then distribute these particles to different ranks/processors. Each processor deals with the same number of particles. The quantities on cell (like field and current) are not parallelized. Each processor sees the same field and current on the whole mesh, pushes particles using the ready fields and easily calculate currents. Advantage: computing load of each rank/processor is absolutely same. Disadvantage: fields are not parallelized. Refer to Section III-F for more information on PIC parallelization

## III. JPIC2d

When one is familiar with the logic of one 1D PIC, he/she can start to write a 2D or 3D PIC code. Only prerequisites are a 2D field solving scheme and current calculation scheme.

### A. Field solving
### 1. P-polarized field

From Eqs. (7,8), one obtains

$$\frac{\partial E_x}{\partial t} = \frac{\partial B_z}{\partial y} - 2\pi J_x \tag{58}$$

$$\frac{\partial E_y}{\partial t} = -\frac{\partial B_z}{\partial x} - 2\pi J_y \tag{59}$$

$$\frac{\partial B_z}{\partial t} = \frac{\partial E_x}{\partial y} - \frac{\partial E_y}{\partial x} \tag{60}$$

Defining $P_r = (E_y + B_z)/2$, $P_l = (E_y - B_z)/2$ same as 1D PIC, Eqs. (59,60) become

$$(\frac{\partial}{\partial t} + \frac{\partial}{\partial x})P_r = \frac{1}{2}\frac{\partial E_x}{\partial y} - \pi J_y \tag{61}$$

$$(\frac{\partial}{\partial t} - \frac{\partial}{\partial x})P_l = -\frac{1}{2}\frac{\partial E_x}{\partial y} - \pi J_y \tag{62}$$

### 2. S-polarized field

From Eqs. (7,8), one obtains

$$\frac{\partial B_x}{\partial t} = -\frac{\partial E_z}{\partial y} \tag{63}$$

$$\frac{\partial B_y}{\partial t} = \frac{\partial E_z}{\partial x} \tag{64}$$

$$\frac{\partial E_z}{\partial t} = \frac{\partial B_y}{\partial x} - \frac{\partial B_x}{\partial y} - 2\pi J_z \tag{65}$$

Defining $S_r = (E_z - B_y)/2$, $S_l = (E_z + B_y)/2$, Eqs. (64,65) become

$$(\frac{\partial}{\partial t} - \frac{\partial}{\partial x})S_l = -\frac{1}{2}\frac{\partial B_x}{\partial y} - \pi J_z \tag{66}$$

$$(\frac{\partial}{\partial t} + \frac{\partial}{\partial x})S_r = -\frac{1}{2}\frac{\partial B_x}{\partial y} - \pi J_z \tag{67}$$

How to solve Eqs. (58,61,62) and Eqs. (63,66,67), please refer to [7]. Same as 1D case, this kind of field solver is a NDFX scheme and needs no any absorption boundary. In this robust scheme, Δt=Δx and Δx<Δy, i.e. lower resolution in y-dir than x-dir for the numerical stability issue. Sentoku already extended it into NFDXY [8], i.e. no numerical dispersion in both x-dir and y-dir. In 3D, it can be extended to a NFDXYZ, i.e. no numerical dispersion along all three axes. Transverse boundary condition for fields is periodic.

**B. Particle pushing**

First, we discuss how to interpolate fields on particle. Some field components are located on cell center and need to be first interpolated onto cell vertice. Linearly interpolate fields (on cell vertice) onto particle (see Fig. 5)

$$\mathbf{E}_p = \mathbf{E}_{i,j}A_{00} + \mathbf{E}_{i+1,j}A_{10} + \mathbf{E}_{i,j+1}A_{01} + \mathbf{E}_{i+1,j+1}A_{11} \tag{68}$$

where the total area/weight is $A_{00} + A_{10} + A_{01} + A_{11} = 1$. Assuming particle position is (x,y), the cell index is $i = \text{int}(x)$, $j = \text{int}(y)$, where (x,y) is already normalized to cell size (Δx,Δy). The weight in Eq. (68) is

$$A_{00} = (i+1-x)(j+1-y) \tag{69-1}$$

$$A_{01} = (i+1-x)(y-j) \tag{69-2}$$

$$A_{10} = (x-i)(j+1-y) \tag{69-3}$$

$$A_{11} = (x-i)(y-j) \tag{69-4}$$

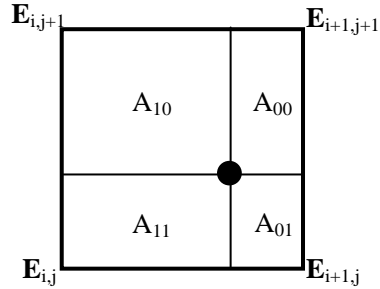

Figure 5. Field is on cell vertice; Particle (black ball) is within the cell.

Pushing particle is carried out by Eq. (33,34,39,45). Y-dir Displacement equation and its finite difference are

$$dy/dt = V_y \tag{70}$$

$$y^{n+1} = y^n + \Delta t V_y^{n+1/2} \tag{71}$$

In real PIC programming, y is normalized to the y-dir cell size Δy. Equation (71) becomes

$$y^{n+1} = y^n + V_y^{n+1/2}\Delta t/\Delta y \tag{72}$$

If some particles move to the x-dir boundary, they are reflected with a thermal velocity like 1D PIC. Y-dir boundary condition for particles is periodic: particles through the top boundary will enter the bottom boundary with the same momentum.

## C. Current and plasma density calculations

We use a new CCS to calculate currents, i.e. zigzag scheme [9]. Please directly refer to Eqs. (14-17) in Ref. [9] (no literal errors) for implementing current calculations. The zigzag scheme can be easily programmed and faster than other CCSs. Author also realizes Esirkepov's CCS [10] in JPIC, and no big difference is found between them. Esirkepov's CCS can be used for any higher-order particle shape. Zigzag scheme is limited to one-order triangle particle shape.

Plasma density defined on cell vertice (only for diagnosis) can be obtained by (see Fig. 5) $n_{i,j} = n_0 A_{00}$, $n_{i,j+1} = n_0 A_{01}$, $n_{i+1,j} = n_0 A_{10}$, and $n_{i+1,j+1} = n_0 A_{11}$, where $n_0$ is given in Eq. (47).

## D. Make Plasma.c

A uniform 2D rectangle plasma slab contains cn_x*cn_y*pn_per_cell particles, where cn_x and cn_y are x-dir and y-dir cell numbers of plasma. One can say that there are $\sqrt{\text{pn\_per\_cell}}$ particles in x-dir or y-dir. Particle separation in x-dir and y-dir can be $ps = 1/\sqrt{\text{pn\_per\_cell}}$. Linear or exponential ramp in x-dir can refer to 1D PIC case. Arbitrary 2D density profile can be generated by a random number generator like 1D PIC.

## E. Make Pulse.c

One should use a 2D Gaussian beam, instead of 3D beam. Following a derivation of 3D Gaussian beam in the textbook, one can obtain the Gaussian beam in 2D space

$$E = E_0 \sqrt{\frac{w_0}{w}} \exp(-\frac{x^2}{w^2}) \exp(i\Phi) \tag{73}$$

$$\Phi = (z/z_R)(x/w)^2 - \phi/2 + kz - \omega t \tag{74}$$

For comparison, 3D beam is

$$E = E_0 \frac{w_0}{w} \exp(-\frac{r^2}{w^2}) \exp(i\Phi), \tag{75}$$

$$\Phi = (z/z_R)(r/w)^2 - \phi + kz - \omega t \tag{76}$$

where beam waist $w = w_0\sqrt{1+(z/z_R)^2}$, $\phi = \arctan(z/z_R)$, and Rayleigh length $z_R = kw_0^2/2$. There are two differences between them: $\sqrt{\frac{w_0}{w}}$ vs. $\frac{w_0}{w}$, and constant coefficient of $\phi$.

## F. Parallelization

Before parallelizing one code, first one should make sure that serial/nonparallel code works well. Second one can learn some basic functions in the MPI package. Since we do not go to details of field solver and current calculation in JPIC2d, here we just give a rough description on parallelization of JPIC2d.

### 1. Parallelization environment

On supercomputers, MPI programming environment is always ready. For convenience, one can set up a MPI

programming environment in the personal computer with Windows system. First, download and install MPICH1 or MPICH2 [11]. Check their manuals and set up Visual Studio 6.0. Then one can write, compile and run a parallel C or C++ code in PC.

## 2. Basic parallelization idea

To calculate $\sum_{i=1}^{1000} a[i]$ in a single CPU, 1000 add operations are carried out. For example, the array is $a[i] = i(i+1)$. The code (pseudo) can be

**For i=1:1000**
**a[i]=i*(i+1);**
**End**
**Sum=0;**
**For i=1:1000**
    **Sum+=a[i];**
**End**

In 10 processors, one can separate 1000 numbers into 10 subgroups and distribute them into these 10 processors. Each processor deals with 100 numbers. The operation process in each processor is called as a rank. For 10 ranks, the rank label is [0:9]. Rank0 is generally the primary rank. The parallel code can be

**For i=1:100**
**ii=100*rank_label+i;**
**a[i]=ii*(ii+1);**
**End**
**Sum=0;**
**For i=1:100**
    **Sum+=a[i];**
**End**
**COLLECT_SUM_RANK0(Sum,0)**
**# 0 means rank0, COLLECT_SUM_RANK0 is not a real MPI function!**

All processors/ranks see the above same code. Use the rank label (from 0 to 9) to initialize different numbers (1:100 on rank0, 101:200 on rank1 …) in the array. Do 100 add operations on different numbers on each rank. Finally, transfer the result Sum in rank1-9 to rank0 and add them together to Sum in rank0. Now Sum in rank0 is the final result. One can use if(rank_label=i) to do any special operation in the ranki. For example: output Sum to screen in rank0. One can use MPI functions to do communications between all the ranks. For example: distribute Sum in rank0 to other ranks and replace Sum there. This simple example contains the main idea on how to parallelize a code.

## 3. JPIC2d parallelization

JPIC2d split the whole 2D simulation space along y-dir into many small equal-size sections. Each processor is responsible to one section, which seems to be a small simulation box and contains its own particles. Neighboring sections share the same border field. Particles shuttles between neighboring sections.

## 3.1 Plasma.c

One can initialize all the particles in the primary rank0 and then distribute them into other ranks according to position y. This particle initialization is limited by the memory of the primary processor. For the simulation with a huge particle number, each rank/section needs initialize its own particles. JPIC2d contains both approaches. Particle's information is stored in linked lists, one kind of data structure. Basic operations on the linked list include: find and delete an element in the list and add/insert an element into the list. They are more complex than an array structure. Advantage of a linked list: dynamic memory allocations according to actually particle number, which is specifically good for new-species generation during PIC simulations, for instance, new-born charged ion species by some ionization processes and electron-positron pair production in vacuum by near Schwinger limit intense laser. Unlike a fixed array, one has to guess possible maximum particle number for one species. Disadvantage: operation on the linked list is complex and slow (memory allocating of a linked list is irregular); for an array, memory is allocated as a continuous block and access is fast. For non-new-species generation PIC, array can be a good choice if one can make a good guess on maximum particle number on each rank during the whole simulation. A bad guess will fail the simulation. Try again with a larger particle number.

## 3.2 Field.c and Push.c

If the total cell number in y-dir is cn_y and total rank/processor number is rank_num, cell number each rank will be cn_y_per_rank=cn_y/rank_num. In the real PIC, each rank has to deal with (cn_y_per_rank+2) cells in y-dir. Additional 2 cells are called guard cells and located on the bottom and top of the section.

In Field.c, for fields on cell center, one needs update these fields on guard cells by use of MPI communication functions so that one can interpolate these boundary fields onto particles near boundaries.

In Push.c, when some particles move out the section and go to guard cells, they will generate currents on guard cells. Then use MPI functions to add currents on guard cells to neighboring sections. Go inside the particle list and kick out particles in both bottom and top guard cells. Store these particles' information into an array and send them to up or bottom neighboring section. In the neighboring section, prepare an empty array for coming particles and add them to the particle list. At first, one need account the number of leaving particles, so that a fit array can be prepared for storing these leaving or coming particles.

Each section can be sending and receiving something (fields and particles). Be careful to avoid message blocks. One can avoid the message block by the means that even ranks first send and then receive, and odd ranks first receive and then send.

## 3.3 Output.c

Try to output data in binary format, which is typically many times smaller than ASCII format and important for massive parallel simulations. In JPIC2d, each rank outputs its own data. For snapshots output (like field $E_x$ on the mesh at one time), one can name the file as ex.time_step.rank_label. If one uses several thousands of processors and even more, the total number of all files can be huge, which will lead to slow reading and post-processing. One may make subfolder like ex/time_step/rank_label, where ex and time_step are folders and rank_label is a file. Parallel output technique can output data from all ranks into a single file. JPIC2d does not use this technique yet.

## 3.4 Are results from parallel JPIC2d same as ones from serial JPIC2d?

Laser pulse propagation in the vacuum can give exactly same results in both serial and parallel JPIC2d, if one is enough careful to make sure that initialization of injecting fields on boundary is exactly same for both serial and parallel codes.

When considering plasma dynamics, both serial and parallel codes can never give the same results. Even in serial code, exchanging the pushing sequence of any two particles will lead to different results. This is due to the fact that add operation of a group of different-sequence floating point numbers will leads to different results. This difference means that noise is different. However, macroscopic physical results should remain same and are independent of the particle pushing sequence.

**3.5 What kinds of MPI functions are used in JPIC2d?**

In Input.c, at the very beginning, use MPI_Init to enter MPI environment, use MPI_Comm_rank to get the label of the current rank, and use MPI_Comm_size to get total number of ranks.

In Field.c, use three functions MPI_Datatype, MPI_Type_vector, MPI_Type_commit to define a special data package of fields, which will be transferred as a whole between different ranks; use MPI_Send and MPI_Recv to send and receive the defined data package. Totally 24 MPI_Send and 24 MPI_Recv functions are used. The numbers of MPI_Send and MPI_Recv are always marched. After MPI communications, use MPI_Type_free to cancel the defined data package.

In Push.c, use MPI_Datatype, MPI_Type_vector, MPI_Type_commit to define three special data packages for Jx,Jy,Jz,; totally 24 MPI_Send and 24 MPI_Recv functions are used to transfer these currents. One additional pair of MPI_Send and MPI_Recv is used to transfer particles passing between neighboring ranks.

In Output.c, use MPI_Barrier to synchronize all ranks, and then use MPI_Reduce to calculate maximum, minimum or sum of same data in all of ranks, finally use MPI_Bcast to let all of ranks to know the results. These operations are used to calculate total energy, maximum or minimum energy/momentum of particles in the whole simulation box.

In Main.c, on the very end, use MPI_Finalize() to close MPI environment.

## IV. Application of JPIC

Author had used JPIC to solve some different problems, which always need more or less changes or modifications on the code. Some examples are shown below, with explanations on both technical points and physics problems. First, some general advices on how to use a PIC code are given.

**A. Using a PIC code**

Making reliable PIC simulations is important. First, user should know the limitation of the code at hand and basic knowledge about used field solver and current calculation algorithm in the code. Whatever commercial, others' or own PIC codes, before using it to do some works, one needs carry out some test runs on several well-known or analyzable problems or do benchmarks with other codes. Simple analyzable problems can be light dispersion relation (group velocity) in underdense plasmas and laser wakefields in the linear regime [12]. PIC can exactly reproduce them. For relativistic laser intensity, laser wakefield is still a good example to make comparison and benchmark, because lost of benchmark simulations using different PIC codes [13] and quasi-static code QuickPIC [14] are already done and modeling is also elaborate and clear [15]. In principle, if there are no big bugs, all PIC codes should give quite similar results, because different field or current calculation algorithms should lead to small difference in most of cases.

For simulating laser underdense plasma interactions, the smallest space and time scales are laser wavelength and laser cycle, which need to be well resolved, like at least 10 grids/cells per wavelength. Cell number per wavelength is various for different cases. Higher intensity laser may need higher resolution. For simulating laser overdense plasma interaction, the smallest space and time scales become plasma wavelength and plasma oscillation period. The rule to choose a proper resolution is making higher and higher resolution simulations until results converge. Convergence means that results stay stable or unchanged when resolution is higher than a critical threshold. Using this critical or higher resolution is safe for further works. Using more particles per cell can decrease the noise level and also have effects on convergence. Strictly speaking, each different problem needs such a convergence test. It should be noted that lower-than-critical-threshold-resolution (LTCTR) simulation is not totally worthless. Sometimes, LTCTR simulation is done due to limited computer sources and can still qualitatively demonstrate some physics phenomena or mechanisms. However, quantitative claims based on LTCTR simulation results may be unreliable.

**B. Oblique incidence and immobile ions & a scheme for single-cycle THz generation**

**1. Technical point**

As discussed above, using a Lorentz transformation in 1D PIC can deal with oblique incidence problems. 1D PIC is applicable when laser spot sizes are much larger than laser wavelength and plasma wavelength. Oblique incidence in 2D is implemented by rotating plasma slabs by the incident angle. For initializing particles per rank in JPIC2d, after rotation, some particles can move from one section to another one and particle transfers by MPI functions are needed. In short pulse underdense plasma interactions, ion motions can be frequently neglected. For immobile ion simulations, one does not need to initialize ions. Initially zero fields imply that there is a virtual ion background to neutralize electrons.

**2. Physics problem**

In Ref. [16], Author proposed a simple scheme to generate a powerful single-cycle THz pulse by obliquely irradiating lasers on an underdense plasma slab. 1D and 2D PIC give almost same results when laser spot size is enough large. THz output scaling with incident angle and detailed radiation mechanism are discussed in Refs. [16] and [17]. MW-GW THz sources can be potentially produced by this scheme, which demands experiment demonstrations.

**C. Atom ionization & phase-sensitive THz generation**

**1. Technical point**

Using linked lists can easily deal with atom ionization in PIC. When an electron or ion is born, they will be added into their linked lists. First, according to local electric field on atom, ionization rate R is calculated. By the same time, call ran() to return a random from 0 to 1. When ran()<R [18], this atom is ionized and will be deleted from its list, and freed electron and ion are added into their lists. ADK rate is a popular formula to calculate the tunneling ionization rate. There is about one-order-magnitude difference between ADK theory and numerical solution of ionization rate for higher intensity laser. However, numerical solutions are only available for simple atoms (H,He). JPIC2d adopts numerical ionization rates. Since few-cycle laser pulses are used to excite THz radiation, Author find that conventional FDTD field solver can not keep the pulse CEP constant even in vacuum. Tens of microns of vacuum propagation will change a sine pulse to a cosine one. This is more serious for shorter pulses. NDFX scheme is used to overcome this problem.

**2. Physics problem**

Stimulated by the experiment work on measuring CEP of few-cycle laser pulses by THz emissions [19], Author added atom ionization process into JPIC2d. In the gas, collision ionization is negligible. In interested laser intensity regime, tunneling ionization is dominant process. In the paper [20], Author clarified CEP dependent THz radiation. JPIC2d gives the same results as analytic model. Especially, circularly polarized few-cycle pulse can manipulate polarization plane of THz pulse by changing its CEP. This supplies a controllable parameter to benefit applications of THz sources. This polarization control demands experiment demonstrations.

**D. Multi-particle-weights & a scheme for uniform relativistic electron sheet generation**

**1. Technical point**

In general, only two species (electron and ion) are allocated in PIC. One can add more species. For a multiple-layer target, electron or ion from each layer can be defined as one species, which can have different weight. For example, matrix X[0:1][0:np] is storing position for electron and ion in the first layer; and X1[0:1][0:np1] can be for the second layer. One can use the same Push.c to push particles with the attention that charge density per macro-particle in each layer can be different. Temperatures in each layer can be different. Respective diagnosis on different species is helpful to distinctly study dynamics of each species.

2. **Physics problem**

In Ref. [21], Author proposed a novel scheme to generate a transverse-momentum-free and uniform relativistic electron sheet (RES) for coherent Thomson scattering (CTS). This method can potentially generate >GW attosecond XUV and x-ray coherent sources. The successful demonstration of such a tool will revolutionize coherent XUV and x-ray sciences in the table-top scale. In the simulation, Author used 5e7 particles in the first foil and several million particles in the second foil. In fact, the second foil is much denser and thicker than the 1st one. The 1st foil is set a 10eV electron temperature and the second foil has no temperature. Plasmas with a certain temperature will emit electromagnetic fields, which may contaminate CTS signals.

**E. Moving window & laser or beam driven wakefield**

**1. Technical point**

When laser or beam interacts with very long plasmas and one is only interested in the region around the driven source, a moving simulation box covering the driven source can be used. This is called as the moving window technique. When right-going driven source is close to the right boundary, simulation box starts to move. For JPIC, there is always $\Delta t=\Delta x$. If the velocity of laser or beam in plasmas is very close to c, after each time step, all physics quantities is shifted by a cell/grid from right to left. The leftmost-cell physics quantities will be replaced and the rightmost ones will be initialized (fields are zero; new fresh particles will be injected). The speed of this special moving window is exactly c. When a driven source has a slow speed (like 0.8c), one can shift 4 cells every 5 time steps ($4\Delta x/5\Delta t=0.8$). When the speed of the driven source is varying, one can set $i\Delta x/j\Delta t$ close to transient source speeds. When driven sources are particle beams (electron, positron or ion), one needs add one additional species for the particle beam. Since there is no light, it is convenient to normalize Maxwell equations to background plasma related units: $\frac{e\mathbf{E}}{m_e\omega_p c}, \frac{e\mathbf{B}}{m_e\omega_p}, \frac{\rho}{en_0}, \frac{\mathbf{J}}{en_0 c}, \frac{\mathbf{r}}{\lambda_p}, \frac{2\pi t}{\omega_p}$, where $n_0$ is background plasma density, $\lambda_p$ and $\omega_p$ is plasma wavelength and frequency of the background plasma. One can find that code does not need any changes. Set background plasma density to unity $n_0=1$. Now, cell number per wavelength in input deck means cell number per plasma wavelength. Beam density $n_b$ is in unit of $n_0$. How to initialize an electron bunch with density $n_b$, energy $\gamma_0$ and energy spread $\Delta\gamma$? First initialize an immobile electron bunch with density $n_b$ (strictly speaking, a plasma bunch is initialized). Set an accelerating length $L_a$ to reach the energy $\gamma_0$. Calculate the needed accelerating field $E_a=-\gamma_0/2\pi L_a$, where all quantities are normalized to the background plasma related units. Set all other fields on beam particle to zero, and only use $E_a$ to accelerate beam. When average energy of bunch reaches $\gamma_0$, set $E_a$ to zero and recover all other fields. Energy spread is realized by accelerating field "spread", i.e. $\Delta E_a=-\Delta\gamma_0/2\pi L_a$. A Gaussian random distribution routine Gaussian() is called and the final accelerating field is $E_a$+Gaussian()*$\Delta E_a$, which will lead to a energy spread $\Delta\gamma$ after the accelerating length $L_a$. Each particle in bunch has a different accelerating field, which needs to be allocated like other information of particle.

**2. Physics problem**

Using JPIC2d with the moving window technique, Ref. [22] shows and validates Tajima's idea on collective deceleration of electron/positron beams by excited wakefields in underdense plasmas, which can serve as a cheap and non-radioactivation beam dump. This could benefit the future high-energy (>GeV) table-top laser wakefield accelerator. We also find a new microbunching mechanism which is induced by nonuniform Betatron oscillation in the beam and explained by a simple model.

**F. Boosted PIC technique & nonlinear CTS for keV coherent attosecond x-ray pulse generation**

**1. Technical point**

For simulating laser/beam wakefield accelerator and free-electron laser (FEL), one can use the simple moving window technique discussed above to significantly speed simulations. This is still not realistic for an extremely long distance multi-dimensional simulation. The boosted PIC technique proposed by Vay [23] can further speed such kind of simulations. The basic principle is that using a Lorentz transformation goes to a moving frame with a relativistic factor $\gamma$, where laser frequency is decreased by the factor $\gamma(1+\beta)$, background plasma density is increased by $\gamma$. PIC simulation is done in this moving frame, just like dealing with oblique incidence in 1D PIC. Compared with PIC in the lab frame, lower resolution due to prolonged laser wavelength and shortened plasma length due to relativistic length contraction largely shrink the simulation efforts and can speed total simulation time by several orders of magnitudes [24]. In addition to making a proper Lorentz transformation, code needs not any changes. This technique also can used to simulate FEL and CTS from RES.

**2. Physics problem**

For a CTS simulation in the lab frame, when the signal is keV x-ray, tens of thousands of cells per laser wavelength are needed for resolve the x-ray wavelength. Author makes a boosted PIC simulation on CTS in the co-moving frame of the RES (see Ref. [25]). In the new frame, RES is immobile and laser and scattered signal have the same frequency. The simulation efforts are much reduced and about two orders of magnitudes speeding is shown. In the analytic theory, Author extended earlier linear CTS results to nonlinear/relativistic CTS with the scattering amplitude $a_{0,s}>1$. Theory and simulation are in good agreement. New nonlinear effects are shown, such as nonlinear Doppler shift factor, output saturation, and pulse contraction and steepening.

**V. Perspective**

General PIC method is pretty mature. There are still broad interests to embed new capabilities into the classical collisionless PIC. For examples: 1. Binary collision can be included to simulate some collisional plasma physics [26],[27]; 2. Radiation damping force can be included to investigate $a_0>100$ laser plasma interactions [28]; 3. Electron-positron pair generation by electromagnetic cascades can be included to study $a_0>1000$ laser vacuum interactions [29]; 4. Using high-order particle shape with lower noise in PIC could be used to investigate dynamics of ultra-cold plasmas [30].

There is another algorithm improvement in PIC: particle positions are normalized to the cell size in a real PIC, and **r**-int(**r**) gives particle displacement in the cell, which determine the precision of current calculations and field interpolations. Since floating point number **r** has finite-number effective digits, the precision of **r**-int(**r**) decreases with the larger cell label, i.e. the whole system has incongruous precision. This problem is more serious for float numbers and first treated in VPIC [31], which adopts two numbers to replace the single **r**. One of them is just the cell label (integer type) and the second is the float displacement in the cell. The precision in the whole simulation box is fixed. The future JPIC may include this approach, which may have slight effects because JPIC uses double variables thoroughly.

Making JPIC3d with NDFX or NDFXYZ field solver and 3D zigzag CCS is on the Author's plan. Generally, 1D and 2D PIC are enough to find and demonstrate new physics mechanisms. 3D PIC is important for certain questions (3D relativistic self-focusing [32] and electron self-injection in bubble/blowout wakefields [33]) and is quite useful to quantitatively compare and predict experiment results.

With the rapid development of GPU high performance computation, parallelizing JPIC1d on GPU is being planned [34]. Typically, 1D PIC demands small memory (<1GB). GPU technique should be perfect for it and 10 times speeding is expected in a single GPU.

**Acknowledgements**

HCW acknowledges support from the Alexander von Humboldt Foundation during 2007-2009, when JPIC2d was developed. HCW wishes to thank the PhD advisor Prof. Z.-M. Sheng for delivering his own 1D PIC to Author and for all the time supports and encouragements when building up JPIC2d. HCW appreciates Prof. Y. Sentoku for his generously sharing his unpublished notes on NDFX and NDFXY. HCW is grateful for Dr. C. Huang, Dr. L. Yin and Dr. B. Albright for their helpful discussions. Thanks kids Jingwen & Junwen for their initial in "JPIC".

**Postscript**

The original purpose to write this paper is for a JPIC reference in Author's future publications. Then Author find out that more details (direct collection of notes at hand) may be beneficial to fresh students or new beginners in the field. Author is open to discuss any details in JPIC1d or JPIC2d and other PIC-related topics. Some opinions in the paper may be subjective and any comments are welcome.

**About Author**

**Hui-Chun Wu** was born in Shanxi, China, on February 7, 1980. He received the Ph.D. degree from the Institute of Physics, Chinese Academy of Sciences, Beijing, China, in 2006. In the same year, he was awarded a Tsai Shi-Dong Plasma Physics Prize, China, and an Alexander von Humboldt Research Fellowship, Germany. He worked as a Postdoctoral Researcher at the MPQ, Germany, during 2007-2009. Currently, he is a Research Associate at the LANL. His research interests are high-field light-matter interactions, especially developments & applications of laser-plasma based coherent photon (THz, X-ray) and particle (electron, ion) sources.